# Total chiral symmetry breaking by bifurcation from racemic systems: when "left" and "right" cannot coexist.

**Cristobal Viedma**

*Departamento de Cristalografia y Mineralogia, Facultad de Geologia, Universidad Complutense, Madrid, Spain. E-mail: viedma@geo.ucm.es*

**Processes that can produce states of broken chiral symmetry are of particular interest to physics, chemistry and biology. Up to now it was believed that total chiral symmetry breaking during crystallization of sodium chlorate occurs via the production of secondary crystals of the same handedness from an initial single chiral crystal or "mother crystal" that seeds the solution. Here we report that a solution of sodium chlorate with a large and "symmetric" population of D- and L-crystals moves into complete chiral purity disappearing one of the enantiomers. This result shows: (i) a new symmetry breaking process, incompatible with the hypothesis of an initial single chiral crystal; (ii) that complete symmetry breaking and total chiral purity can be achieved from an initial system with both enantiomers. These findings demand a new explanation to the process of total symmetry breaking in crystallization without the intervention of an initial single chiral phase and open the debate on this fascinating phenomenon. We describe the autocatalytic process that connected with Ostwald ripening enhances any small initial crystal enantiomeric excess and gives total symmetry breaking.**

*Introduction*. – Chiral symmetry breaking occurs when a physical or chemical process that does not have preference for the production of one or other enantiomer spontaneously generates a large excess of one of the two enantiomers: (L), left-handed or (D), right handed . From the energetic point of view, these two enantiomers can exist with an equal probability and inorganic processes involving chiral products commonly yield a racemic mixture of both (L) and (D) enantiomers [1]. However life on earth utilizes only one type of amino acids and only one type of natural sugars: (L)-amino acids and (D)-sugars. The fact that biologically relevant molecules exist only as one of the two enantiomers is a fascinating example of complete symmetry breaking in chirality and has long intrigued many scientists. With a few exceptions the symmetry breaking produced by different natural mechanisms have proved giving small enantiomeric excess (EE) [2], ranging from the 20% found experimentally for asymmetric photolysis, to the $10^{-17}$ alleged theoretically for parity violating energy difference between enantiomers. This means that to reach total chiral purity, mechanisms to enhance any initial imbalance in chirality are absolutely essential [3].

In 1953 Frank [4] suggested that a form of autocatalysis in which each enantiomer catalyses its own production, while suppressing that of its mirror image, might have nonlinear dynamics leading to the amplification of small initial fluctuations in the concentrations of the enantiomers. Many theoretical models are proposed afterwards, but they are often criticized as lacking any experimental support [5]. Recently, asymmetric autocatalysis of pyrimidyl alkanol has been studied intensively [6] [7] [8] [9]. The enhancement of EE was confirmed [7], and its temporal evolution was explained by the second-order autocatalytic reaction [8] [9]. But only with the nonlinear autocatalysis, chirality selection is not complete and the value of EE stays less than 100%.

In a recent theoretical model Saito [10] confirms that in a closed system, the nonlinear autocatalysis amplifies the initial small enantiomeric excess but eventually, the simple back reaction can promote the decomposition of less abundant enantiomer to the reactant, which is recycled to produce the more abundant type. Through this recycling process, the complete chiral purity can be achieved.

In this Letter we show the first experimental case where total symmetry breaking is achieved from a system where both enantiomers are present since the beginning.

Firstly, we show with laboratory experiments how an isothermal saturated solution of sodium chlorate ($NaClO_3$) with a large population of D- and L-crystals moves into complete chiral purity : (i) any small initial crystal enantiomeric excess (CEE) eventually gives rise to total crystal purity disappearing the less abundant enantiomer (100% CEE); (ii) "symmetric" proportion of both enantiomeric crystals gives rise to total symmetry breaking and crystal purity disappearing randomly one of the two enantiomers. We study the key factors that promote this behaviour. Secondly, up to now it was believed that any case of total symmetry breaking during crystallization occurs via the production of secondary crystals of the same handedness from a single "mother crystal" that seeds the solution. However, our results show a case of total symmetry breaking incompatible with the latter idea, i.e., we are dealing with an entirely different process.

Furthermore the experimental results show that complete symmetry breaking and chiral purity can be achieved from an initial system with both enantiomers. Therefore the findings demand a new explanation for this process of total symmetry breaking and open the debate on this fascinating phenomenon. Finally, we present arguments indicating that Ostwald ripening process in a solution of sodium chlorate is sufficient to give rise to total symmetry breaking and chiral purity from a previously racemic medium.

*Sodium chlorate crystallization.-* The achiral molecules of $NaClO_3$ crystallize as two enantiomeric chiral crystals in the cubic space group $P2_13$ [11]. Hence sodium chlorate is achiral before crystallization, as it exists in solution as more or less dissociated ions or clusters without a fixed chirality, but forms a chiral crystal,

About 100 years ago Kipping and Pope [12] demonstrated that solutions of $NaClO_3$ by seeding can produce total CEE. More recently Kondepudi and others [13] showed that simply stirring during the crystallization of sodium chlorate from solution was sufficient to produce a yield approaching 100% of just one enantiomer. Whereas under normal conditions a distribution of the proportion of one or other enantiomer obtained in a series of experiments falls on a typical Gaussian curve, with the peak yield at 50% of each enantiomer, in the stirred experiments the distribution is bimodal, with the peak yield close to 100% of one or the other enantiomer. Kondepudi suggested that the most important factor in this chiral symmetry breaking is secondary nucleation by which a seed crystal or randomly generated "mother" crystal triggers the production of a large number of secondary crystals at a fast rate if the solution is stirred that are enantiomerically identical to itself. The result of this crystallization process is the generation of crystals with the same handedness (total chiral symmetry breaking). Nevertheless, the hypothesis of an initial single chiral crystal to explain spontaneous chiral symmetry breaking has been disputed recently [14].

*Experimental.* - Chiral L- and D-crystals were obtained separately by literature procedures via seeding with the appropriate chiral crystal a supersaturated solution [15]. We prepared samples (12 g) with mixtures of L- and D-$NaClO_3$ chiral crystals with 5% CEE (L-crystals: D-Crystals = 5.7000 g : 6.3000 g or 6.3000 g : 5.7000 gr.), and samples with "symmetric" mixtures of both crystals (L-crystals: D-crystals = 6.0000 g : 6.0000 g). Every sample was ground using an agate pestle and a mortar and was placed in a set of 25 mL round-bottom flasks with 10 mL of water. Because of solubility of $NaClO_3$ [16], an excess of more than 2 g. of crystals remains without dissolving at 24ºC. We added to some solution flasks, 8 g, 6g, 4g, or 2g of small glass balls (3 mm of diameter), that continuously crushed the crystals that were being subjected to stirring by a magnetic bar (3-20 mm) at 600 rpm. Flasks with only solutions and crystals and flasks with solutions, crystals and glass balls were hermetically closed and maintained at 24ºC with constant agitation. Every solution contains a population of more than a million of crystals. The continuous mechanical abrasion of crystals by glass balls imposes to crystals a maximum size of about 200 μm that is kept during all the experiment although they grow (it depends fundamentally on the size of the glass balls because the system works like a mill). A parallel experiment consisted in 4 flasks with solution, crystals and 4 g of glass balls every one with constant agitation but with different speed of agitation ( 200, 400, 600 and 800 rpm).

Samples of solution with crystals (0.1 mL with more than 10,000 crystals) were removed from the flasks every 2 hours and left to grow during a few hours. The experiments are considered finished when chiral crystal purity is achieved. Chiral crystal evolution was determined by their optical activity using a petrographic microscope. Due to the optical isotropy of the crystals, light passing through the crystals in every direction exhibits optical activity in the same sense; there is no need to orient the crystals in any particular direction to observe optical activity. Under transmission, a D crystal will rotate light counter clockwise while a L crystal will rotate light clockwise. Therefore when crystals are placed between crossed polarizers they will appear darker or lighter than the other species when one of the polarizers is uncrossed by a few degrees (Fig.1).

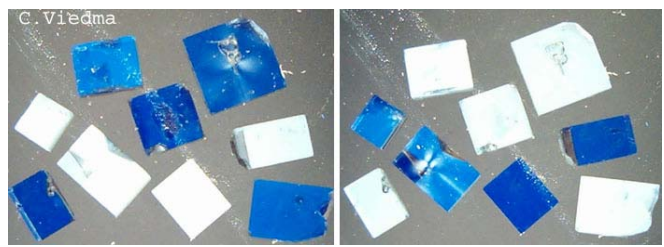

FIG 1 Left: L crystals show blue colour and D crystals white when we rotate the polarizer a few degrees clockwise. Right: L crystals show white colour and D crystals blue when we rotate the polarizer a few degrees counter clockwise.

The ease with which the L- and D-crystals could be identified enables us to know when chiral crystal purity has been reached. All experiments have been performed and repeated more than 20 times in successive weeks. The reported results are average values.

*Results and discussion.* - Crystals in the stirred solutions (no glass balls involved) maintain indefinably (several days) the initial enantiomeric excess or the initial "symmetry" between both populations of L- and D-crystals. However, all stirred crystals-solution systems where glass balls (abrasion-grinding process) intervene show a continuously enhancement of CEE and eventually, total symmetry breaking and complete chiral crystal purity is achieved at different times (Fig. 2).

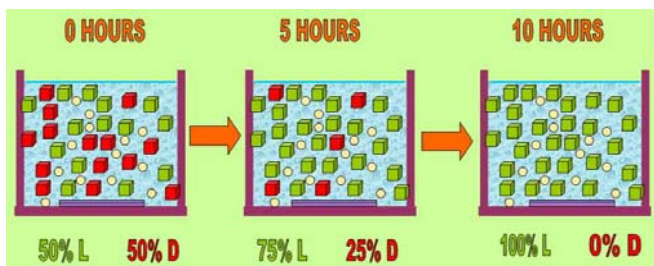

FIG. 2 Chiral crystal purity is achieved in a few hours from an initial racemic system.

The time required to achieve chiral purity depends on the number of glass balls present in the system (Fig. 3), and when the number of glass ball is the same it depends on the speed of agitation of the system (rpm) (Fig.4).

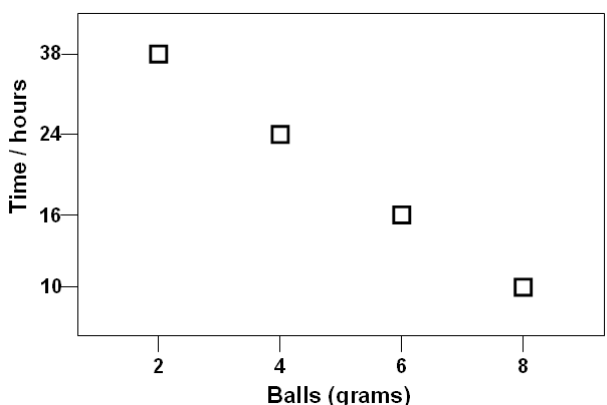

FIG 3. Solutions with initial "symmetric" mixtures of L and D-crystals and glass balls show total symmetry breaking and chiral crystal purity. The data show the necessary time to achieve chiral purity depending on number of balls in the system (600 rpm).

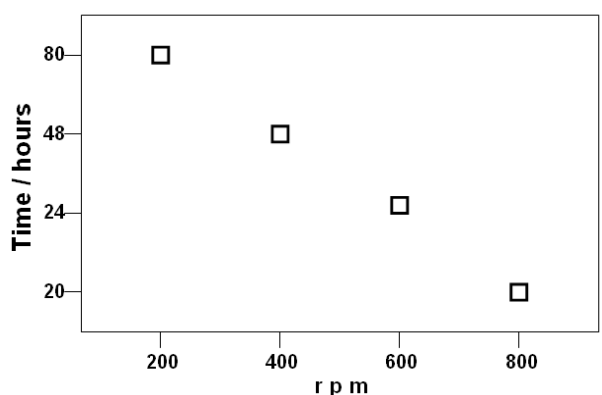

FIG 4 Solutions with initial "symmetric" mixtures of L and D-crystals and glass balls show total symmetry breaking and chiral crystal purity. The data show the necessary time to achieve chiral purity depending on the speed of agitation of the system (4gr of balls).

After 8 hours, solutions with initial 5% L-CEE show 100% L-CEE, and solutions with initial 5% D-CEE show 100% D-CEE (Fig.5).

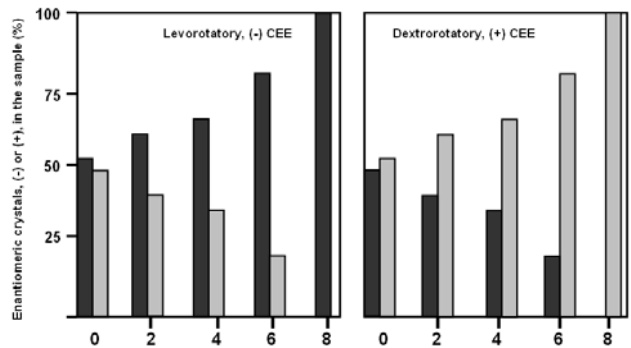

FIG. 5 Asymmetric evolution of chiral crystals. Solutions with initial 5% L-CEE show 100% L-CEE, and solutions with initial 5% D-CEE show 100% D-CEE in 8 hours (600 rpm).

Solutions with initial "symmetric" mixtures of L and D-crystals and 4 gr of glass balls show total symmetry breaking and chiral purity after 24 hours (Fig. 6). The handedness in this last case is L or D (randomly). Nevertheless, any small difference between L and D-crystals induces the preferred production of one of them, for example small differences in the quality of the crystals bias the progressive enantiomeric amplification of a certain handedness.

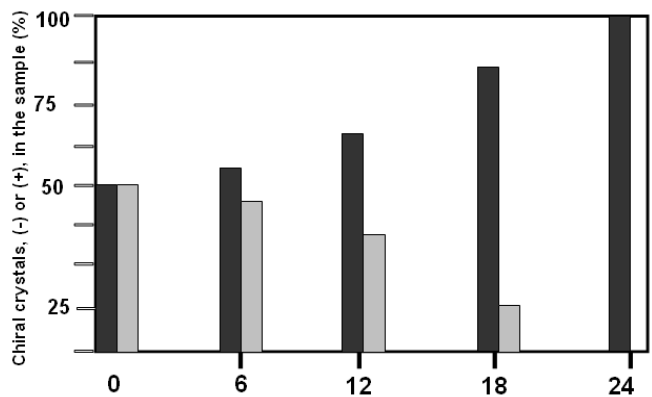

FIG 6 Initial "symmetric" mixtures of D and L-crystals and 4 gr of balls show total symmetry breaking and chiral purity after 24 hours (600 rpm). The handedness in this case is L or D randomly.

It is evident from these results that under these conditions complete chiral purity cannot be explained by the model of an initial single chiral crystal, and therefore, a new explanation is required. Simple statistic tells us that at small scale the most likely state is one in which one enantiomer dominates slightly the other [17]. Thus, we need a mechanism that amplifies this slight initial enantiomeric excess to total chiral purity.

The crystal enantiomeric excess can be measured by CEE = $(N_L - N_D) / (N_L + N_D)$, where $N_L$ and $N_D$ are the number of each enantiomer, that is a means to quantify the symmetry breaking.

Secondary nucleation is the nucleation of new crystals caused by the presence of an existing primary crystal. Recently Cartwright [18] established with theoretical argument that mechanical crushing and detaching asperities from the surface of the primary crystal by supplying sufficient force to rupture are on the microscale the same mechanism that for secondary nucleation. In fact collisions between one crystal an another, or between a crystal and the fluid boundaries (container walls, stirring bar etc) break pieces off the surface, and fluid shear, which may also detach fragments from the crystal surface, have each been put forward as responsible for homochiral secondary nucleation [19] [20]. The detached fragment will possess the same chirality as the crystal of which they previously formed a part.

Thus we can consider that the abrasion-grinding process by the glass balls of our experiments is a sort of "induced" secondary nucleation that generates new fragments of crystals with the same chirality that the mother crystal.

Following the idea of Frank that an nonlinear autocatalysis amplifies the small initial fluctuations in the concentrations of the enantiomers, Cartwright and coworkers [18] show with numerical simulations that Secondary nucleation can act as just such a nonlinear autocatalytic process, as, in the case of chiral crystallization, secondary nuclei possess the same chirality as the mother crystal, so the presence of a crystal of a given chirality catalyses the production of further crystals with the same chirality.

To demonstrate these ideas, they construct a minimal model for chirality selection via secondary nucleation in solution by building on the work on nonlinear autocatalytic processes in flows of Metcalfe and Ottino. At first, there is no enantiomeric excess, but by the end, they observe an overwhelming predominance of one enantiomer, corresponding to a value of CEE of 0:899. They show with these simulations that secondary nucleation is a nonlinear autocatalytic process capable of explaining the earlier experimental results of symmetry breaking in crystallization.

However, in their simulation we can observe again that only with the nonlinear autocatalysis, chirality selection is not complete and the value of CEE stays less than unity. They assume that to obtain chiral purity we need "a single mother crystal: an ancestral Eve for the whole population".

These same results are achieved by Saito and Hyuga [10] with a mathematical model: they consider a chemical reaction such that substances A and B react to form substance C. Though reactants A and B are achiral, the product C happened to be chiral in two enantiometric forms; R-isomer (R)-C and S-isomer (S)-C.

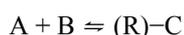

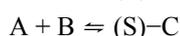

They found that the EE amplification takes place with a nonlinear autocatalytic chemical reaction. However, the final chirality is not complete: the EE is smaller than 100%. However the inclusion of the back reaction from the products (R)-C and (S)-C to A and B, brings about the drastic change in the results. The component which has a slight advantage starts to dominate, and the other chiral type extinguishes gradually. Eventually, the complete homochirality or chiral purity is achieved.

It is to say the nonlinear autocatalytic process amplifies a small initial EE and afterwards the simple back reaction promotes the decomposition of the less abundant enantiomer to the reactant, which is recycled to produce the more abundant enantiomer. Through this recycling process, the complete chiral purity can be achieved.

Due to the combined abrasion-grinding (glass balls) and stirring in our experiments, the left and right handed crystals continuously loose tiny fragments of left and right hand micro-crystallites or clusters; it is to say we are dealing with an "induced" secondary nucleation and because of this, with a nonlinear autocatalytic phenomenon. But… where is the recycling process?

We are dealing with a population of crystals in contact with its solution, and therefore, the surface of crystals is subjected a continuous dissolution-crystallization process. Since the abrasion-grinding process results in the continuous generation of micro-crystals the effective surface area of crystalline phase continuously increase. Additionally, the thermodynamic prediction of Gibbs-Thompson equation states that "solubility" of the crystalline phase in a solution is dependent on particle size [21]. One of the principal ageing process for a precipitate that remains in contact with its mother liquor is Ostwald ripening. It is a process at chemical equilibrium that leads to dissolve smaller particles feeding the larger ones. Thus in the abrasion-grinding process of our system the finest fraction of micro-crystallites or clusters easily dissolve, feeding the larger ones. That is to say, we are dealing with a continuous dissolution-crystallization phenomenon or Ostwald ripening only that in this case the process is highly enhanced by the abrasion-grinding process.

Since the chirality of Sodium chlorate is a property of its crystal structure, any molecular arrangement greater than the unit cell is chiral. For $NaClO_3$ Z (number of molecules in unit cell) = 4 [22]. Thus any molecular group of more than 4 molecules generated from a crystal is chiral and any molecular group of less than 4 molecules has not crystal structure and reaches the achiral molecular level.

During the dissolution process the final stage of any crystallite or chiral cluster is at this achiral molecular level, and therefore can feed other crystals independently of its chirality (the majority enantiomer has more advantage). Thus, in this process the chiral crystals are continuously recycled (Fig. 7).

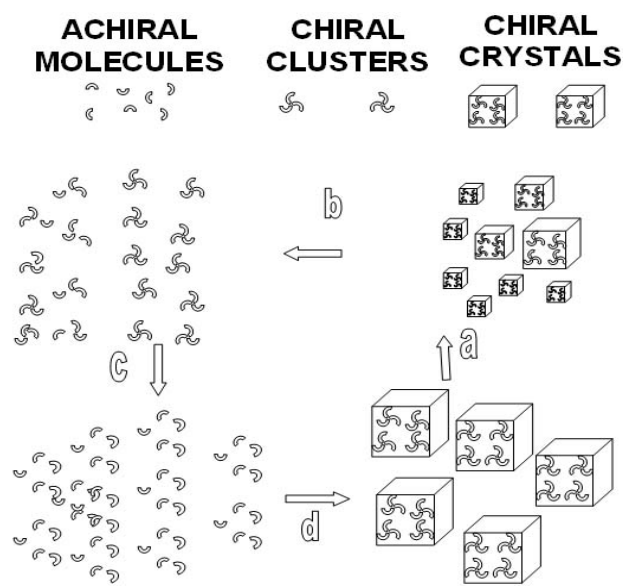

FIG 7. Recycling process. The abrasion-grinding process generates micro-crystals that easily dissolve (a). The final stage of any crystallite or chiral cluster is the achiral molecular level (b) (c). These molecules feed other crystals independently of its chirality (d).

Very recently Uwaha (23), knowing these experimental results, constructs a simple mathematical model based on a few physical assumptions that mimic our experimental system. The development of this model shows complete chiral symmetry breaking in a crystallization process. He agrees with us in that the recycling process occurs when a crystal reaches the achiral molecular level. He concludes that the symmetry breaking and realization of homochirality is purely a dynamical effect. In his model the dissolution-crystallization process is spontaneous and the interchange of matter induced by differences in particles size is not considered. Nevertheless we believe that both mechanisms of dissolution can work simultaneously in our specific experiments.

Whatever the exact mechanism, it is clear from our experiments that complete fast homochirality cannot be explained by the traditional theory of secondary nucleation from an initial single chiral crystal. Therefore there must be other mechanisms that amplify any initial fluctuation in the system to achieve symmetry breaking and complete chiral purity.

In this respect, we may speculate that a solution of sodium chlorate, strongly agitated in an isothermal slow evaporation process, can undergo similar pathways at the microscopic scale (micro-crystals). At this level the effective surface of solid phase is maximum, resulting in total symmetry breaking that is manifested at the macroscopic level. Additionally, chiral clusters smaller than the critical nucleus size may also be subjected to this symmetry breaking process. In turn, when these clusters reach the critical nucleus size, a chiral symmetry breaking in crystallization from primary nucleation will "inherit" the chiral condition achieved during the pre-nucleation stage. These hypotheses must be regarded as complementary explanations to the phenomenon of symmetry breaking described in other works [13], [14].

*Conclusion.-*

We show experimental data indicating that complete homochirality and chiral purity can be achieved from an initial system where both enantiomers are present: this is the first experimental case in which one observes the complete elimination of a chiral population of a hand in favour of the other one.

We establish that a common process of ageing, such as Ostwald ripening, in a selective and autocatalytic dissolution-crystallization process can overwhelm the slightest initial chiral imbalance.

In our system this becomes possible by the combination of nonlinear autocatalytic dynamic of the abrasion-grinding and the recycling of crystallites when they reach the achiral molecular level in the dissolution-crystallization process.

In this respect we propose that chiral purity and total symmetry breaking during crystallization can be achieved without the necessary intervention of an initial single chiral phase or "mother crystal".

Beyond the specific aspects of these experiments, a significant fact can be established with far reaching implications: final and total chiral purity seems to be an inexorable exigency in the course of some natural processes.